\documentclass[preprint,amsmath,amssymb,superscriptaddress,floatfix,nofootinbib,11pt]{revtex4-1}
\usepackage[utf8]{inputenc}
\usepackage{mathrsfs,bm}
\usepackage{natbib}
\usepackage{txfonts}
\usepackage{amssymb}
\usepackage{rotating}
\usepackage{indentfirst}
\usepackage{graphicx,booktabs}
\usepackage{multirow}
\usepackage{slashed}
\usepackage{overpic}
\usepackage{color}
\usepackage{amssymb}
\usepackage{hyperref}
\usepackage{bbding}
\usepackage{url}
\usepackage{ulem}

\begin{document}

\title{Discriminating 1D new physics solutions in $b\to s\ell\ell$ decays}

\author{Shuang-Yi Li}
\affiliation{School of Physics,  Beihang University, Beijing 102206, China}

\author{Rui-Xiang Shi}
\affiliation{School of Physics,  Beihang University, Beijing 102206, China}

\author{Li-Sheng Geng}
\email{E-mail:lisheng.geng@buaa.edu.cn}
\affiliation{School of Physics \& Beijing Key Laboratory of Advanced Nuclear Materials and Physics, Beihang  University, Beijing 102206, China}
 \affiliation{School of Physics and Microelectronics,
Zhengzhou University, Zhengzhou, Henan 450001, China}

\begin{abstract}
The recent measurements of $R_{K^+}$, $R_{K_S^0}$, $R_{K^{*+}}$, $B_s\to\mu^+\mu^-$, a set of CP-averaged angular observables for the $B^0\to K^{*0}\mu^+\mu^-$ decay, and its isospin partner $B^+\to K^{*+}\mu^+\mu^-$  by the LHCb Collaboration, consistently hint at lepton universality violation in the $b\to s\ell\ell$ transitions. In this work, we first perform global fits  to the $b\to s\ell\ell$ data and show that five one-dimensional scenarios, i.e, $\delta C_9^{\mu}$, $\delta C_{10}^{\mu}$, $\delta C_L^{\mu}$, $\delta C_9^{\mu}=C_{10}^{\mu\prime}$, and $\delta C_9^{\mu}=-C_9^{\mu\prime}$ can best explain the so-called B anamolies. Furthermore, we explore how these scenarios can be distinguished from each other. For this purpose, we first study the combinations of four angular asymmetries $A_i$~$(i=3,4,5,9)$ and find that they cannot distinguish the five new physics scenarios. We then show that a newly constructed ratio $R_{S}$ can uniquely discriminate the five new physics scenarios in proper intervals of $q^2$ if it can be measured with  a percent level precision.
\end{abstract}

\maketitle

\section{Introduction}

In the standard model~(SM), the rare decays of $B$ mesons induced by flavour-changing neutral-current~(FCNC) $b\to s\ell\ell$ transition 
 at the quark level are suppressed by the GIM mechanism~\cite{Glashow:1970gm}. Therefore they provide an ideal laboratory to indirectly probe new physics (NP) beyond the SM. Interestingly,  the  measurements of several observables yield results in tension with the SM expectations (see Refs.~\cite{Li:2018lxi,Bifani:2018zmi} for recent reviews).
 
 Recently, the LHCb Collaboration reported the most precise measurement of $R_{K^+}=\Gamma(B^+\to K^+\mu^+\mu^-)/\Gamma(B^+\to K^+e^+e^-)$ in the bin $[1.1,6]~{\rm GeV}^2$~\cite{LHCb:2021trn}
\begin{eqnarray}
R_{K^+}=0.846^{+0.042 +0.013}_{-0.039 -0.012},
\end{eqnarray}
showing that the significance of deviation from the SM prediction (see Table 1 in Ref.~\cite{Geng:2017svp}) is at  $3.1\sigma$ confidence level. Compared to the 2014 measurement~\cite{Aaij:2014ora}, the tension with respect to the SM prediction has significantly increased. At the same time, a new result for the $B_s\to\mu^+\mu^-$ branching fraction,
\begin{eqnarray}
{\rm BR}(B^0_s\to\mu^+\mu^-)=(3.09^{+0.46 +0.15}_{-0.43 -0.11})\times10^{-9},
\end{eqnarray}
is also published by the LHCb Collaboration~\cite{LHCb:2021awg,LHCb:2021vsc}, which allows for a better constraint on the Wilson coefficient $C_{10}$. In Ref.~\cite{Geng:2021nhg}, together with the theoretical prediction of Ref.~\cite{Beneke:2019slt} including the double-logarithmic QED and QCD corrections, the CMS measurement~\cite{Sirunyan:2019xdu}, and the  ATLAS measurement~\cite{Aaboud:2018mst}, the following ratio is obtained 
\begin{eqnarray}
R=\frac{{\rm BR}(B_s^0\to\mu^+\mu^-)_{\rm exp}}{{\rm BR}(B_s^0\to\mu^+\mu^-)_{\rm SM}}=0.78(9).
\end{eqnarray}

Very recently, the LHCb Collaboration measured two new lepton-universality ratios $R_{K_S^0}=\Gamma(B^0\to K_S^0\mu^+\mu^-)/\Gamma(B^0\to K_S^0 e^+ e^-)$ and $R_{K^{*+}}=\Gamma(B^+\to K^{*+}\mu^+\mu^-)/\Gamma(B^+\to K^{*+} e^+ e^-)$ in the $q^2$ ranges $[1.1,6.0]~{\rm GeV}^2$ and $[0.045,6.0]~{\rm GeV}^2$, respectively, using proton-proton collision data corresponding to an integrated luminosity of 9 fb$^{-1}$~\cite{LHCb:2021lvy}. The two ratios are
\begin{eqnarray}
&&R_{K_S^0}=0.66_{~-0.14~-0.04}^{~+0.20~+0.02},\nonumber\\
&&R_{K^{*+}}=0.70_{~-0.13~-0.04}^{~+0.18~+0.03},
\end{eqnarray}
where the first error is statistical and the second is systematic. We note that the central values are close to the Belle measurements~\cite{Belle:2019oag,Abdesselam:2019lab}, but the uncertainties are smaller, resulting in deviations from  the SM predictions with a significance of $1.5\sigma$ and $1.4\sigma$, respectively. 

On the other hand, in 2020 the LHCb Collaboration reported angular analyses of the $B^0\to K^{*0}\mu^+\mu^-$ decay and its isospin partner $B^+\to K^{*+}\mu^+\mu^-$ decay~\cite{Aaij:2020nrf,Aaij:2020ruw}. The results reconfirm the global tension with respect to the SM predictions previously reported for the decay of $B^0\to K^{*0}\mu^+\mu^-$~\cite{Aaij:2015oid}.


These new measurements have attracted much attention and led to many model-independent global  analyses~\cite{Altmannshofer:2021qrr,Cornella:2021sby,Kriewald:2021hfc,Alguero:2021anc,Hurth:2021nsi,Geng:2021nhg,Carvunis:2021jga,Angelescu:2021lln,Isidori:2021vtc,Alda:2021krg,Isidori:2021tzd,Lee:2021ldv,Bause:2021ply,Hurth:2021nsi}  assuming the presence of NP only in the $b\to s\mu^+\mu^-$ mode, and it is shown that the significance  of  the  SM  exclusion  in  the  global  fits  is  about $4\sim6\sigma$. These data have also been explained in NP models involving tree-level
exchanges of new particles such as the neutral gauge boson Z'~\cite{Cen:2021iwv,Kawamura:2021ygg,Allanach:2021gmj,Alok:2021pdh,Ko:2021lpx,Navarro:2021sfb,Bause:2021prv,Alok:2021ydy,Wang:2021uqz,Davighi:2021oel}, leptoquarks~\cite{Angelescu:2021lln,Greljo:2021xmg,Cornella:2021sby,Kriewald:2021hfc,Nomura:2021oeu,Du:2021zkq,Perez:2021ddi,Greljo:2021npi,King:2021jeo,Singirala:2021gok,Chang:2021axw,Perez:2021ddi,Ban:2021tos,Marzocca:2021azj,Lee:2021jdr} or scalar Higgs~\cite{Chen:2021vzk,Duan:2021whx,Chen:2021vzk}. It is interesting to note that the leptoquark models can simultaneously explain the $b\to s\ell^+\ell^-$ (with $\ell=e,\mu$) and charged current~(CC) $b\to c\ell^-\bar{\nu}_\ell$ (with $\ell=\mu,\tau$) flavor anomalies~\cite{Babu:2020hun,Angelescu:2021lln,Cornella:2021sby,Kriewald:2021hfc,Du:2021zkq,King:2021jeo,Ban:2021tos}.

The so-called $B$ anamolies can be best explained in five one-dimensional scenarios, i.e, $\delta C_9^{\mu}$, $\delta C_{10}^{\mu}$, $\delta C_L^{\mu}$, $\delta C_9^{\mu}=C_{10}^{\mu\prime}$, and $\delta C_9^{\mu}=-C_9^{\mu\prime}$, as demonstrated in the model independent analyses~\cite{Alok:2019ufo,Alguero:2021anc,Geng:2021nhg,Altmannshofer:2021qrr,MunirBhutta:2020ber,Alok:2020mvm,Kumar:2019qbv,Alok:2017jgr,Capdevila:2017bsm,Alguero:2021yus,Hurth:2021nsi,Hurth:2017hxg,Capdevila:2016ivx,Altmannshofer:2017yso,Descotes-Genon:2013wba,Descotes-Genon:2015uva,Angelescu:2021lln,Altmannshofer:2021qrr,Cornella:2021sby,Kriewald:2021hfc,Isidori:2021vtc,Lee:2021ldv,Hurth:2021nsi}. In this work, assuming that NP appears in the muon mode~\footnote{For a discussion about  the scenario where NP appears in both the electron and muon channels, see Appendix A.}, we first present  global fits to the $b\to s\ell^+\ell^-$ data for the SM and five one-dimensional NP scenarios $\delta C_9^{\mu}$, $\delta C_{10}^{\mu}$, $\delta C_L^{\mu}$, $\delta C_9^{\mu}=C_{10}^{\mu\prime}$, and $\delta C_9^{\mu}=-C_9^{\mu\prime}$, the latter two were not considered in Ref.~\cite{Geng:2021nhg}. Next, we study the four angular asymmetries $A_i$~($i=3,4,5,9$)  and find that they cannot discriminate the five new physics scenarios. Finally, we show that a set of ratios $R_i$ are helpful to uniquely discriminate the NP scenarios. In addition, we present predictions in the SM and different NP scenarios for binned observables $A_i$ and $R_i$ in certain ranges of bins, which might be relevant for future experiments. 

This work is organized as follows. We introduce the observables $A_i$ and $R_i$ in Sec.~\ref{Sec2}. Results and discussions are given in Sec.~\ref{Sec3}, followed by a short summary and outlook in Sec.~\ref{Sec4}.

\section{Angular observables $A_i$ and $R_i$ for the $B\to K^*\ell^+\ell^-$ decay}\label{Sec2}

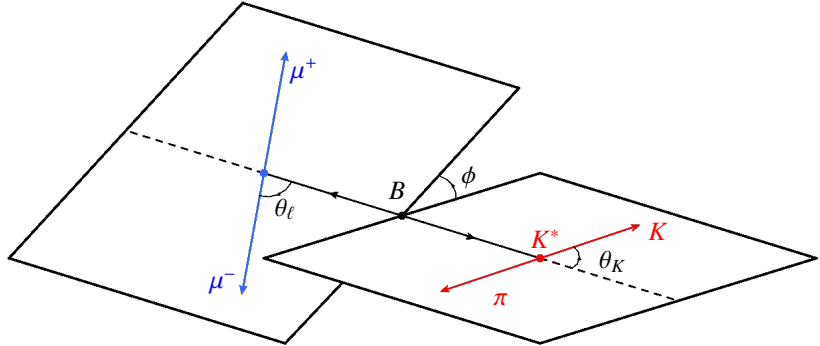
\begin{figure}[!h]
\begin{overpic}[scale=0.25]{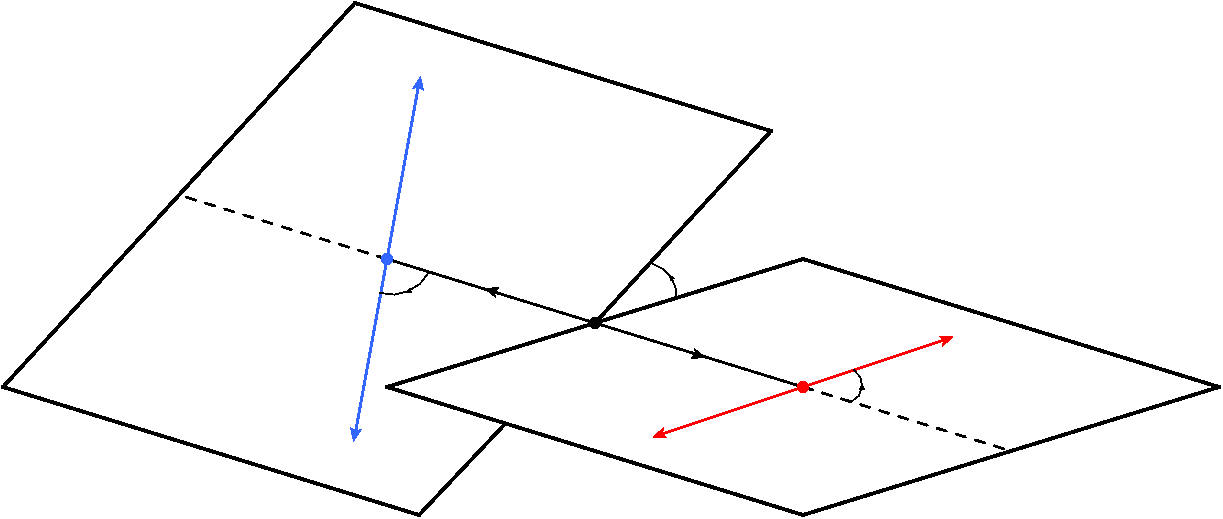}
\put(35,33){$\color{blue}\mu^+$}
\put(25,7){$\color{blue}\mu^-$}
\put(33,16){$\theta_{\ell}$}
\put(47,18){$B$}
\put(56,20){$\phi$}
\put(64.5,12){$\color{red}K^*$}
\put(79,13){$\color{red}K$}
\put(60,5){$\color{red}\pi$}
\put(73,9.5){$\theta_K$}
\end{overpic}
  \caption{Kinematics of the $B\to K^*\ell^+\ell^-$ decay.}\label{Fig1:kinematics}
\end{figure}

For details about our theoretical framework, the low-energy effective Hamiltonian, the origin of theoretical uncertainties and how they are parameterized, and the statistical methods used to perform global fits to all the relevant experimental data, we refer to Ref.~\cite{Geng:2021nhg}.
 In this section, we introduce some angular observables which can potentially discriminate different NP scenarios. For this, we focus on the $B\to K^*\ell^+\ell^-$ decay.  The differential decay rate for the four-body $B(p)\to K^*(k)(\to K\pi)\ell^+(q_1)\ell^-(q_2)$ decay is of the following form~\cite{Altmannshofer:2008dz,Jager:2012uw}
\begin{eqnarray}
&&\frac{d^{(4)}\Gamma}{dq^2 d\cos\theta_{\ell}d\cos\theta_K d\phi} = \frac{9}{32\pi}I^{(\ell)}(q^2,\theta_\ell,\theta_K,\phi),
\end{eqnarray}
with
\begin{eqnarray}
I^{(\ell)}(q^2,\theta_\ell,\theta_K,\phi)=&&\left(I_1^s\sin^2\theta_K + I_1^c\cos^2\theta_K + (I_2^s\sin^2\theta_K + I_2^c\cos^2\theta_K)\cos2\theta_{\ell}\right.  \nonumber\\
&&+ I_3\sin^2\theta_K\sin^2 \theta_{\ell}\cos2\phi+ I_4\sin2\theta_K\sin2\theta_{\ell}\cos\phi\nonumber\\
&&+ I_5\sin2\theta_K\sin\theta_{\ell}\cos\phi\nonumber\\
&&+ (I_6^s\sin^2\theta_K + I_6^c\cos^2\theta_K)\cos\theta_{\ell} + I_7\sin2\theta_K\sin\theta_{\ell}\sin\phi\nonumber\\
&&+\left. I_8\sin2\theta_K\sin2\theta_{\ell}\sin\phi+ I_9\sin^2\theta_K\sin^2 \theta_{\ell}\sin2\phi\right),\label{Eq:angC}
\end{eqnarray}
where the kinematic variables $q^2$, $\theta_{\ell}$, $\theta_K$ and $\phi$ are defined, respectively, as follows: (i) $q^2=(p-k)^2$ is the square of the dilepton invariant mass, (ii) $\theta_{\ell}$ is the angle  between the flight direction of the $B$ meson and the $\ell^-$ lepton in the dilepton rest frame, (iii) $\theta_K$ is the angle between momenta of the $B$ meson and the $K$  meson in the dimeson~($K\pi$) rest frame,\textcolor{red}{} (iv) $\phi$ is the angle between the dimeson~($K\pi$) and dilepton rest frames. As shown in Fig.~\ref{Fig1:kinematics}, both $\theta_{\ell}$ and $\theta_K$ are defined in the interval~$[0,\pi]$ while the range of $\phi$ is $[0,2\pi]$. The angular coefficients $I_i$ expressed in terms of the helicity amplitudes can be found in Ref.~\cite{Jager:2012uw}, while the expressions of the CP-conjugate decay $\bar{B}\to\bar{K}^*\ell^+\ell^-$ can be obtained by the following replacements
\begin{align}
I_{1s(c),2s(c),3,4,7}\rightarrow \bar{I}_{1s(c),2s(c),3,4,7},\qquad I_{5,6,8,9}\rightarrow-\bar{I}_{5,6,8,9},
\end{align}
where  $\bar{I}_i=I_i^{*}$.

Following the strategy outlined in Refs.~\cite{Mandal:2014kma,Alok:2020bia}, we can ``construct'' four angular asymmetries $A_i$~($i=3,4,5,9$) by integrating over different intervals in the angles $\theta_\ell$, $\theta_K$, and $\phi$ for the $B\to K^*\mu^+\mu^-$ decay as follows:
\begin{eqnarray}
&&A_3(q^2)=\frac{\left[\int_{-\pi/4}^{\pi/4}-\int_{\pi/4}^{3\pi/4}+\int_{-\pi/4}^{3\pi/4}-\int_{3\pi/4}^{7\pi/4}\right]d\phi
\int_{-1}^{1}d\cos\theta_K\int_{-1}^{1}d\cos\theta_{\ell}\cdot I_{\rm sum}(q^2,\theta_\ell,\theta_K,\phi)}
{\int_{0}^{2\pi}d\phi\int_{-1}^{1}d\cos\theta_K\int_{-1}^{1}d\cos\theta_{\ell}\cdot I_{\rm sum}(q^2,\theta_\ell,\theta_K,\phi)},\nonumber\\
&&A_4(q^2)=\frac{\left[\int_{0}^{\pi/2}-\int_{\pi/2}^{\pi}+\int_{0}^{\pi}-\int_{\pi}^{2\pi}\right]d\phi
\left[\int_{0}^{1}-\int_{-1}^{0}\right]d\cos\theta_K\left[\int_{0}^{1}-\int_{-1}^{0}\right]d\cos\theta_{\ell}\cdot I_{\rm sum}(q^2,\theta_\ell,\theta_K,\phi)}
{\int_{0}^{2\pi}d\phi\int_{-1}^{1}d\cos\theta_K\int_{-1}^{1}d\cos\theta_{\ell}\cdot I_{\rm sum}(q^2,\theta_\ell,\theta_K,\phi)},\nonumber\\
&&A_5(q^2)=\frac{\left[\int_{0}^{\pi/2}-\int_{\pi/2}^{\pi}+\int_{0}^{\pi}-\int_{\pi}^{2\pi}\right]d\phi
\left[\int_{0}^{1}-\int_{-1}^{0}\right]d\cos\theta_K\int_{-1}^{1}d\cos\theta_{\ell}\cdot I_{\rm diff}(q^2,\theta_\ell,\theta_K,\phi)}
{\int_{0}^{2\pi}d\phi\int_{-1}^{1}d\cos\theta_K\int_{-1}^{1}d\cos\theta_{\ell}\cdot I_{\rm sum}(q^2,\theta_\ell,\theta_K,\phi)},\nonumber\\
&&A_9(q^2)=\frac{\left[\int_{0}^{\pi/2}-\int_{\pi/2}^{\pi}+\int_{0}^{\pi}-\int_{\pi}^{2\pi}\right]d\phi
\int_{-1}^{1}d\cos\theta_K\int_{-1}^{1}d\cos\theta_{\ell}\cdot I_{\rm diff}(q^2,\theta_\ell,\theta_K,\phi)}
{\int_{0}^{2\pi}d\phi\int_{-1}^{1}d\cos\theta_K\int_{-1}^{1}d\cos\theta_{\ell}\cdot I_{\rm sum}(q^2,\theta_\ell,\theta_K,\phi)},
\end{eqnarray}
with
\begin{eqnarray}
&&I_{\rm sum}(q^2,\theta_\ell,\theta_K,\phi)=I^{(\ell)}(q^2,\theta_\ell,\theta_K,\phi)+\bar{I}^{(\ell)}(q^2,\theta_\ell,\theta_K,\phi),\nonumber\\
&&I_{\rm diff}(q^2,\theta_\ell,\theta_K,\phi)=I^{(\ell)}(q^2,\theta_\ell,\theta_K,\phi)-\bar{I}^{(\ell)}(q^2,\theta_\ell,\theta_K,\phi),
\end{eqnarray}
where the angular asymmetries $A_i$ are related to the CP-averaged observables $S_i$,~\footnote{The CP-averaged observables are introduced in Ref.~\cite{Altmannshofer:2008dz} as
\begin{eqnarray}
S_i(q^2)=\frac{I_i(q^2)+\bar{I}_i(q^2)}{d\left(\Gamma+\bar{\Gamma}\right)/dq^2}.
\end{eqnarray}
} namely,
\begin{eqnarray}
&&A_3=\frac{1}{\pi}S_3,\qquad A_4=\frac{1}{\pi}S_4,\nonumber\\
&&A_5=\frac{3}{8}S_5,\qquad A_9=\frac{1}{\pi}S_9.
\end{eqnarray}
 Clearly, analogous to  $S_i$~\cite{Altmannshofer:2008dz,Hurth:2016fbr,Bhattacharya:2019dot}, the observables $A_i$ are also sensitive to NP. 
 As a matter of fact, the relative uncertainties of observables $A_i$ and those of $S_i$ are equal. On the experimental side, these observables can be measured  either by fitting to angular distributions, or by performing  angular integrations of $I_i$. However, this may result in different statistical uncertainties because of different angular coverages. It should be noted that the observables $A_i$ are ratios of combinations of the well-known angular coefficients $I_i$. Therefore it is feasible to measure them in the current experiments.

In addition to the $A_i$ observables, following  Refs.~\cite{Jager:2014rwa,Geng:2017svp}, we revisit the ratios of angular observables $R_i$,
\begin{eqnarray}
R_i(q^2)=\frac{I_i^{(\mu)}(q^2)+\bar{I}_i^{(\mu)}(q^2)}{I_i^{(e)}(q^2)+\bar{I}_i^{(e)}(q^2)},
\end{eqnarray}
which are defined with the angular coefficients in Eq.~(\ref{Eq:angC}). As pointed out in Ref.~\cite{Jager:2014rwa}, the observables $R_i$ are sensitive to new physics and their hadronic uncertainties are almost exactly cancelled. It should be emphasized that in this work the lepton mass $m_\ell$ is non-vanishing and the angular coefficient $I_6^c$ is absent because we donot consider scalar operators.

\section{Results and discussions}\label{Sec3}
In this section, we first update the global fits of Ref.~\cite{Geng:2021nhg} by considering the latest measurements $R_{K_S^0}$ and $R_{K^{*+}}$~\cite{LHCb:2021lvy} and perform  global fits for two more NP scenarios, in comparison with Ref.~\cite{Geng:2021nhg}. Next, we study whether with the four asymmetry observables $A_i$ and the  ratios $R_i$ introduced in Refs.~\cite{Jager:2014rwa,Geng:2017svp}, one can distinguish different one-dimensional NP scenarios.

\subsection{Updated global fits}
\begin{figure}[h!]
  \includegraphics[width=8.5cm]{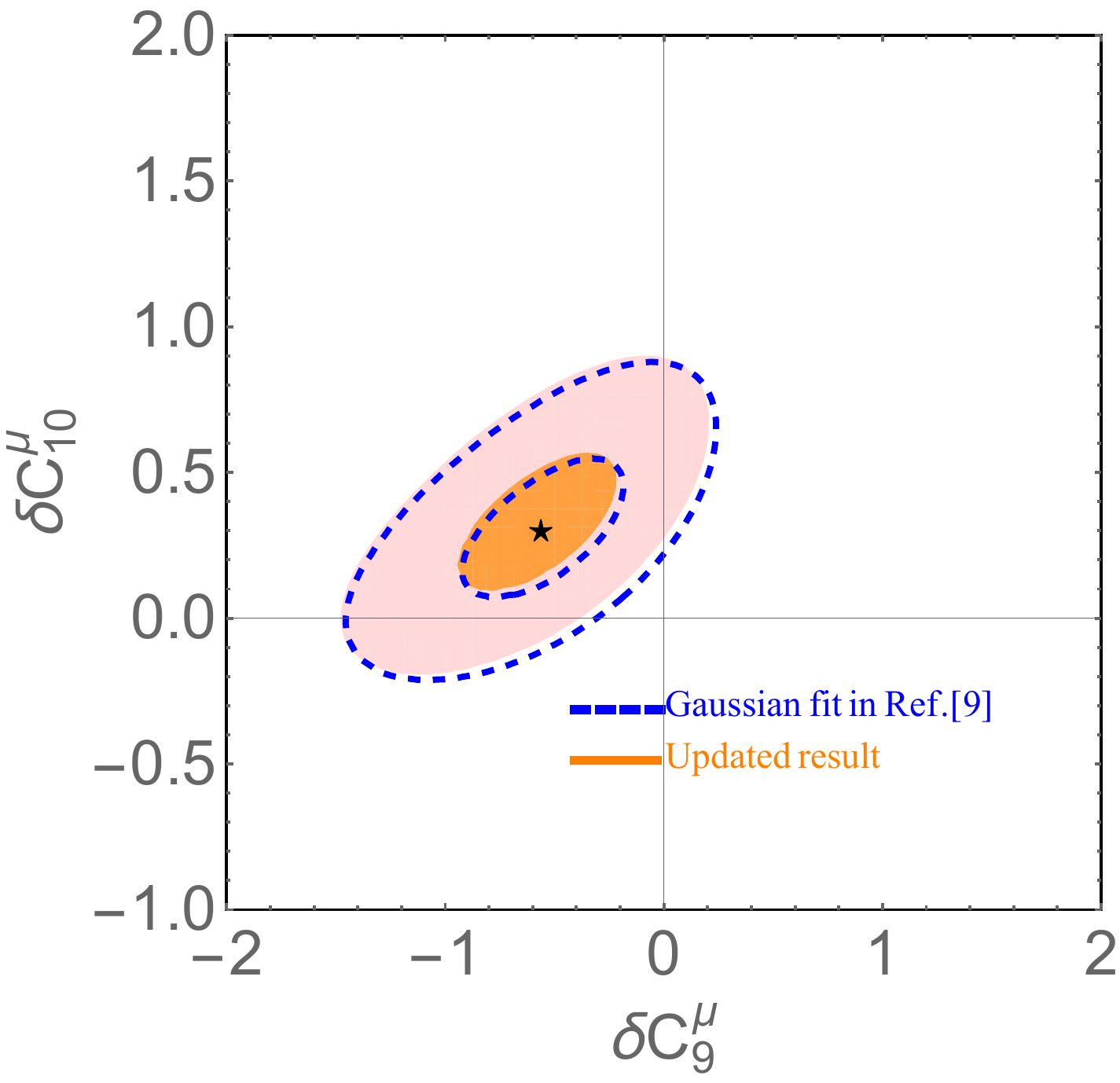}\\
  \caption{Contour plots at the $1\sigma$ and $3\sigma$ confidence level for the $(\delta C_9^{\mu},\delta C_{10}^{\mu})$ scenario.  The updated global fit with all observables for Gaussian~(regions in light red and orange) is compared with the corresponding global fit of Ref.~\cite{Geng:2021nhg}~(dashed lines in blue).}\label{globalfitsPlot}
\end{figure}

In the Appendix of Ref.~\cite{Geng:2021nhg}, we have updated the clean fits insensitive to hadronic uncertainties by considering  the latest measurements  of $R_{K_S^0}$ and $R_{K^{*+}}$. It is shown that the three one-dimensional scenarios $\delta C_9^{\mu}$, $\delta C_{10}^{\mu}$ and $\delta C_L^{\mu}=\delta C_9^{\mu}=-\delta C_{10}^{\mu}$ and a two-dimensional scenario $(\delta C_9^\mu,\delta C_{10}^\mu)$ can well describe the current experimental data. Interestingly, a likelihood ratio test favours the $\delta C_L^\mu$ scenario over the SM at $5\sigma$ significance. 
On the other hand, although the angular observables of the $B\to K^*\mu^+\mu^-$ decay suffer from large hadronic uncertainties, they can still provide constraints on new physics. Therefore, it is interesting to study how the conclusions of the clean fits change when including these angular observables. For this purpose, we update the global fits  of Ref.~\cite{Geng:2021nhg}. We should note that the total number of fitted data for the updated global fits is 96. We obtain $\chi_{\rm SM,min}^2=131.98$ with 96 degrees of freedom, corresponding to a $p$-value of $0.009$. Compared to the global fits of Ref.~\cite{Geng:2021nhg}, the updated fits given in Table~\ref{globalfits} and Fig.~\ref{globalfitsPlot} show that the confidence level of the exclusion of the SM predictions increases by $0.3\sigma\sim0.38\sigma$ for all the NP scenarios. In particular, the significance of deviation from SM in the $\delta C_L^\mu$ scenario is more than $5\sigma$ as well. In addition, we find that the updated global fit  constrains better $\delta C_9^\mu$  in the two-dimensional $(\delta C_9^\mu,\delta C_{10}^\mu)$ scenario and excludes positive values at $1\sigma$ confidence level, as shown in Fig.~\ref{globalfitsPlot}.

In fact, in the case of NP involving tree-level
exchanges of new particles such as the neutral gauge boson Z', leptoquarks or scalar Higgs, further correlations can be induced between the Wilson coefficients $C_9^{\mu(\prime)}$ and $C_{10}^{\mu(\prime)}$. 
As demonstrated in some model independent analyses~\cite{Alok:2019ufo,Alguero:2021anc,Kumar:2019qbv,Alok:2017jgr,Capdevila:2017bsm,Capdevila:2016ivx,Descotes-Genon:2015uva}, two additional one-dimensional scenarios $\delta C_9^{\mu}= C_{10}^{\mu\prime}$ and $\delta C_9^{\mu}=- C_9^{\mu\prime}$ cannot be completely excluded by the current $b\to s\ell\ell$ data. The former scenario can be realized in the $U_1$ leptoquark model in the case of  $\texttt{g}_{\ell q}^{\mu b}\left(\texttt{g}_{\ell q}^{\mu s}\right)^*=\texttt{g}_{ed}^{\mu b}\left(\texttt{g}_{ed}^{\mu s}\right)^*$~\cite{Du:2021zkq}. The latter one could be realized in $Z'$ models with vector-like fermions and $L_\mu-L_\tau$ symmetry ~\cite{Altmannshofer:2014cfa}. Therefore, we would like to investigate these two NP scenarios in our theoretical framework. This can be done following the same strategy as that of Ref.~\cite{Geng:2021nhg} and the corresponding results  are shown  in  Table~\ref{globalfits}. We note that the $\delta C_9^{\mu}=C_{10}^{\mu\prime}$ and $\delta C_9^{\mu}=-C_9^{\mu\prime}$ scenarios can well explain the new $b\to s\ell\ell$ data and the  deviation from the SM has a significance of more than $3\sigma$, smaller than the three one-dimensional $\delta C_9^{\mu}$, $\delta C_{10}^{\mu}$ and $\delta C_L^{\mu}$ scenarios. Indeed, we also tested the other one-dimensional scenarios induced by NP models with the new data studied in the present work and found that they are excluded at  $3\sigma$ confidence level, consistent with the conclusions of Refs.~\cite{Alok:2019ufo,Alguero:2021anc,Kumar:2019qbv,Alok:2017jgr,Capdevila:2017bsm,Capdevila:2016ivx,Descotes-Genon:2015uva}.
\begin{table}[h!]
    \centering
\caption{Best fit values, $\chi_{\rm min}^2$, $p$-value, ${\rm Pull}_{\rm SM}$ and confidence intervals of the WCs in the five one dimensional scenarios.}
\label{globalfits}
{\normalsize
  \begin{tabular}{ccccccc}
\hline
\hline
Coeff. & Best fit & $\chi^2_{\rm min}$ & ~~~$P$-value~~~ & ~~~${\rm Pull}_{\rm SM}$~~~ & 1$\sigma$ range & 3$\sigma$ range\\\cline{1-5}
\hline
$\delta C_9^{\mu}=C_{10}^{\mu\prime}$ & -0.49 & 118.75 [95 dof] & 0.05 & 3.64 & [-0.62, -0.35] & [-0.90, -0.09] \\
$\delta C_9^{\mu}=-C_9^{\mu\prime}$ & -1.03 & 117.99 [95 dof] & 0.06 & 3.74 & [-1.31,-0.76] & [-1.91, -0.20] \\
$\delta C_9^{\mu}$& -0.90 & 108.05 [95 dof] & 0.17 & 4.89 & [-1.11, -0.70] & [-1.55, -0.33] \\
$\delta C_{10}^{\mu}$ & 0.57 & 109.64 [95 dof] & 0.14 & 4.73 & [0.45, 0.70] & [0.20, 0.98] \\
$\delta C_L^{\mu}$ & $-0.42$ & 104.00 [95 dof] & 0.25 & 5.29 & $[-0.50, -0.34]$ & $[-0.68, -0.18]$ \\
\hline
$(\delta C_9^{\mu},\delta C_{10}^{\rm \mu})$ & $(-0.58, 0.32)$ & $103.51$~[94 d.o.f.] & $0.24$ & $4.97$
   & $\delta C_9^\mu \in$ $[-0.81, -0.34]$ & $\delta C_{10}^\mu \in$ $[0.17, 0.48]$\\
\hline
\hline
\end{tabular}
}
\end{table}

As discussed above, among the new NP scenarios studied, all of the five one-dimensional cases, i.e., $\delta C_9^{\mu}$, $\delta C_{10}^{\mu}$, $\delta C_L^{\mu}$, $\delta C_9^{\mu}= C_{10}^{\mu\prime}$ and $\delta C_9^{\mu}=-C_9^{\mu\prime}$, can provide a good description of the latest data. Thus, a natural question is which observables can discriminate between them. For such a purpose, we study the four angular asymmetries $A_i$~($i=3,4,5,9$) and the ratios $R_i$ introduced in Sec. II.

\subsection{Sensitivity of angular asymmetries $A_i$ to new physics}

\begin{figure}[h!]
\begin{tabular}{cc}
  \includegraphics[width=7cm,height=4.5cm]{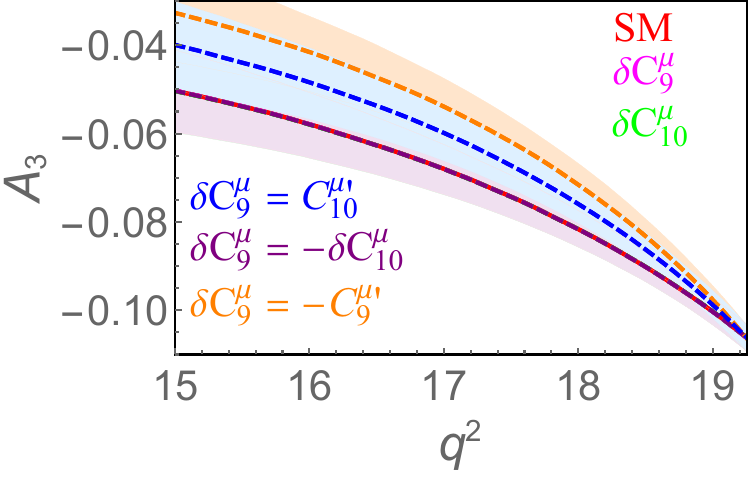}~~~\includegraphics[width=7cm,height=4.5cm]{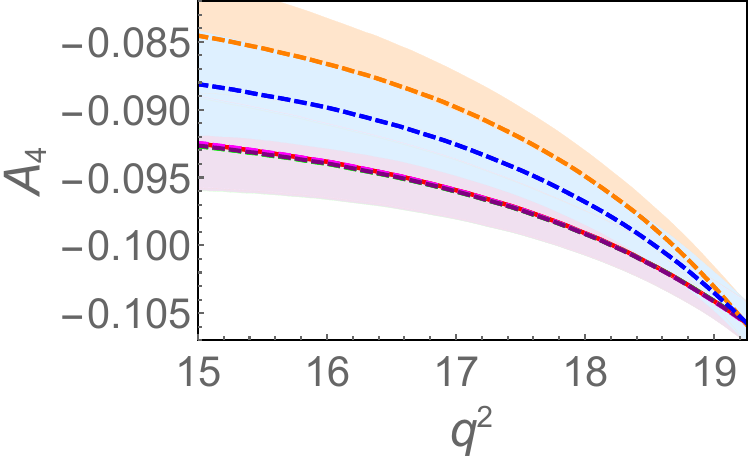}\\
  \includegraphics[width=7cm,height=4.5cm]{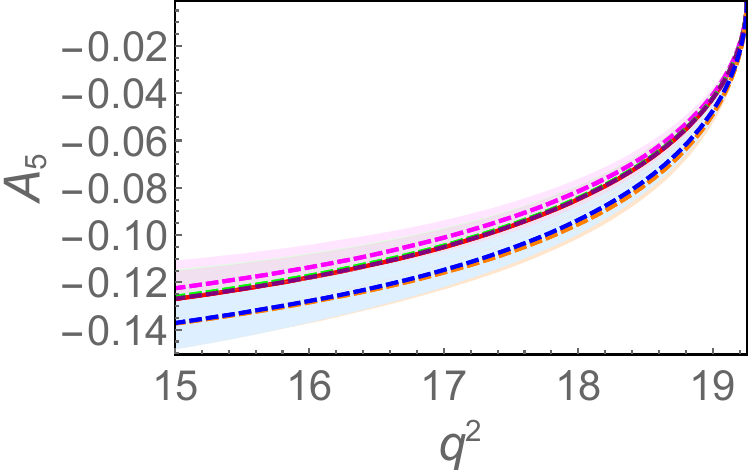}~~~\includegraphics[width=7cm,height=4.5cm]{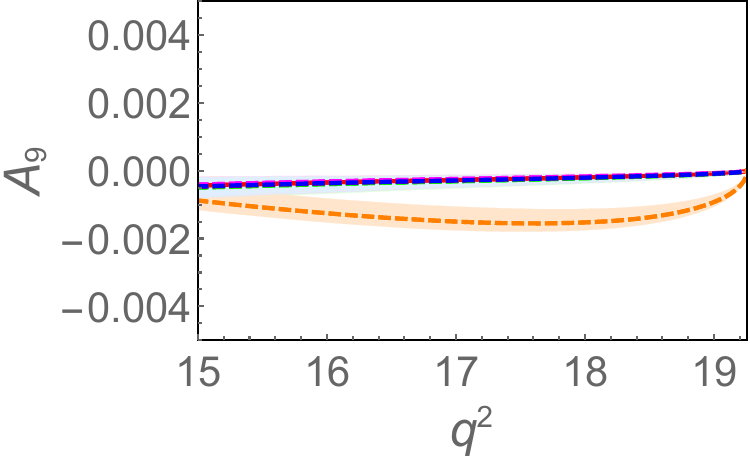}
\end{tabular}
\caption{Dilepton mass squared ($q^2$) spectra of observables $A_i$  in the SM~(solid lines in red) and different NP scenarios~(dashed lines). Shaded regions denote the uncertainties
of SM and five NP scenarios.}\label{Fig:Obsq2}
\end{figure}

In Fig.~\ref{Fig:Obsq2}, we plot $A_{3,5,4,9}$ as functions of $q^2$, where $q$ is the dilepton invariant mass, with the best fitted Wilson coefficients determined for the SM and the five one-dimensional NP scenarios. The shaded bands reflect the uncertainties originating from form factors and contributions of charm loops~\cite{Jager:2012uw,Jager:2014rwa,Geng:2017svp}. The asymmetries integrated over [15,19] GeV$^2$ are given in Table~\ref{Tab:BinnedObs}.
\begin{table}[h!]
\centering
\caption{Predictions in the SM and five NP scenarios for binned observables.}\label{Tab:BinnedObs}
\begin{tabular}{ccccccccc}
\\ \hline \hline
~~~Observables~~~ & ~~~SM~~~ &~~~$\delta C_9^{\mu}$~~~ & ~~~$\delta C_{10}^{\mu}$~~~ & ~~~$\delta C_9^{\mu}= C_{10}^{\mu\prime}$~~~ & ~~~$\delta C_L^{\mu}$~~~& ~~~$\delta C_9^{\mu}=- C_9^{\mu\prime}$~~~ \\ \hline

$A_3$ [15,19] GeV$^2$ & $-0.067^{+0.006}_{-0.006}$ & $-0.067^{+0.006}_{-0.006}$ & $-0.067^{+0.006}_{-0.006}$ & $-0.060^{+0.007}_{-0.007}$ & $-0.067^{+0.006}_{-0.006}$&  $-0.054^{+0.007}_{-0.007}$  \\

$A_4$ [15,19] GeV$^2$ & $-0.096^{+0.002}_{-0.002}$ & $-0.096^{+0.002}_{-0.002}$ & $-0.096^{+0.002}_{-0.002}$ & $-0.093^{+0.003}_{-0.002}$ & $-0.096^{+0.002}_{-0.002}$ &  $-0.090^{+0.003}_{-0.003}$ \\

$A_5$ [15,19] GeV$^2$ & $-0.104^{+0.008}_{-0.008}$ & $-0.101^{+0.008}_{-0.008}$ & $-0.104^{+0.008}_{-0.008}$ & $-0.113^{+0.008}_{-0.008}$ & $-0.104^{+0.008}_{-0.008}$ & $-0.115^{+0.008}_{-0.008}$  \\

$A_9$ [15,19] GeV$^2$ & $-0.0003^{+0.0001}_{-0.0002}$ & $-0.0003^{+0.0001}_{-0.0002}$ & $-0.0003^{+0.0002}_{-0.0002}$ & $-0.0003^{+0.0002}_{-0.0002}$ & $-0.0003^{+0.0002}_{-0.0002}$ & $-0.0013^{+0.0004}_{-0.0002}$
 \\ \hline \hline
\end{tabular}
\end{table}

By examining Fig.~\ref{Fig:Obsq2} and Table~\ref{Tab:BinnedObs}, we can see that these observables can provide two possible strategies to distinguish between different NP structures. The first strategy is to analyse the $q^2$ spectra of certain observables.  It is clear that the discriminating power of these observables $A_i(q^2)$ is very limited. For instance, to distinguish the NP scenario $\delta C_9^{\mu}=- C_9^{\mu\prime}$, an experimental uncertainty of sub percent level or lower is needed. The second strategy is to study binned observables in proper ranges of $q^2$. As shown in Table~\ref{Tab:BinnedObs}, only binned $A_3$ and $A_9$ have the ability to discriminate the scenarios $\delta C_9^{\mu}=-C_9^{\mu\prime}$ and $\delta C_9^{\mu}=C_{10}^{\mu\prime}$ at  $3\sigma$ confidence level if they can be measured with an uncertainty of 0.001. However, the other NP scenarios cannot be discriminated by these observables $A_i$.

In conclusion, the angular asymmetries $A_i$ cannot uniquely distinguish between the five different NP solutions in the one-dimensional scenario.

\subsection{Sensitivity of ratios $R_i$ to new physics}
In this subsection, we study whether the ratios $R_i$ introduced in Refs.~\cite{Jager:2014rwa,Geng:2017svp} can be used to discriminate the five NP scenarios. 

\begin{figure}[h!]
\begin{tabular}{cc}
  \includegraphics[width=7cm]{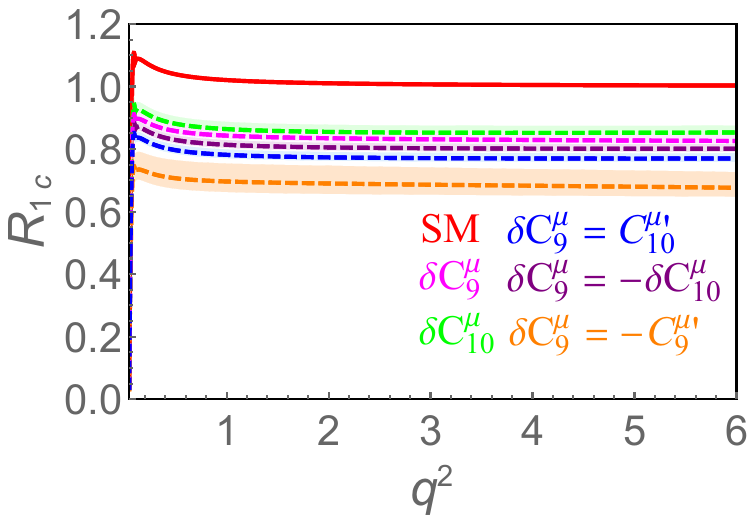}~~~\includegraphics[width=7cm]{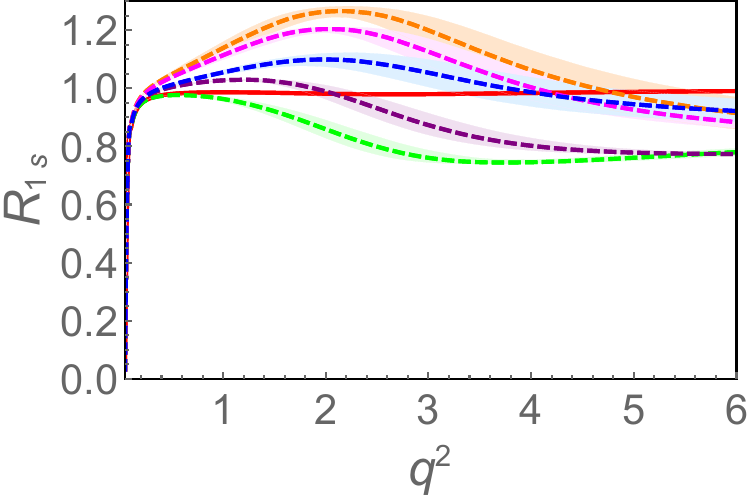}\\
  \includegraphics[width=7cm]{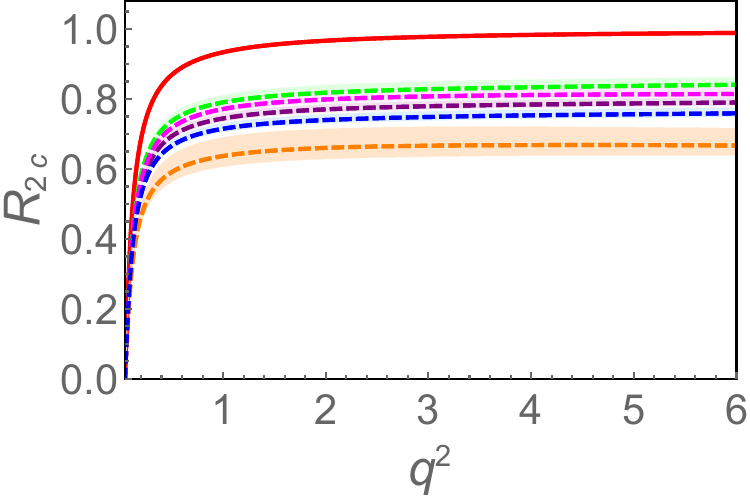}~~~\includegraphics[width=7cm]{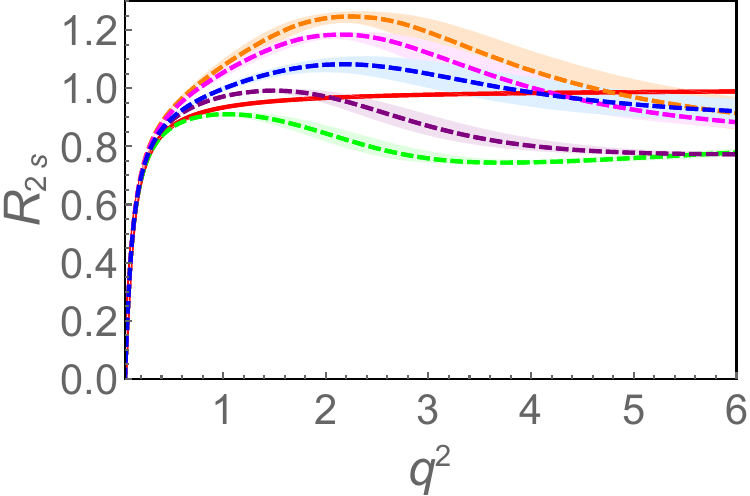}\\
  \includegraphics[width=7cm]{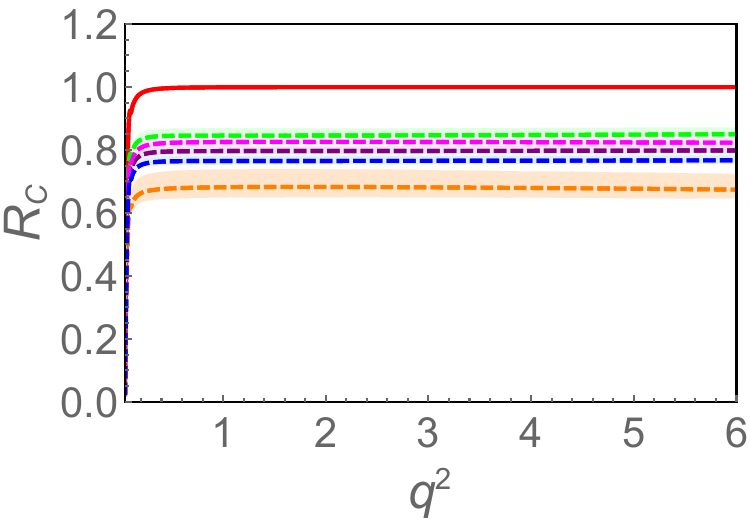}~~~\includegraphics[width=7cm]{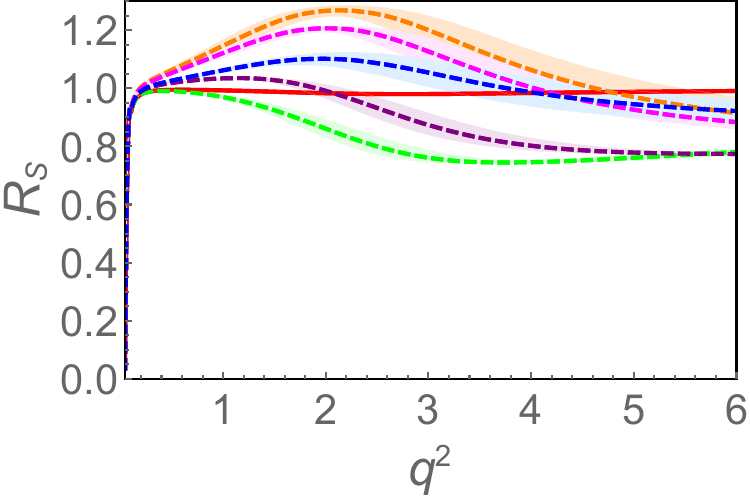}
\end{tabular}
\caption{Same as Fig.~\ref{Fig:Obsq2} but for observables $R_{1c,1s,2c,2s,C,S}$.}\label{Fig:Obsq2Ri}
\end{figure}

First we plot the $q^2$ spectra of $R_i(q^2)$ in the low bin of [0.045,6] GeV$^2$ in Fig.~\ref{Fig:Obsq2Ri}. One can see that the two observable $R_{1s}(q^2)$ and $R_{2s}(q^2)$ could discriminate all the five different NP scenarios, while the  observable $R_{1c}(q^2)$ or $R_{2c}(q^2)$ can only discriminate the NP scenario $\delta C_9^{\mu}=- C_9^{\mu\prime}$. As explicitly shown in Appendix B, the other ratios suffer from uncertainties induced by  power corrections and charm loop contributions and have no ability to distinguish the NP scenarios. We stress that it is possible to distinguish the five different NP scenarios if future experimental statistics is high enough such that an uncertainty of percent level or lower can be achieved.
\begin{table}[h!]
\centering
\caption{Same as Table.~\ref{Tab:BinnedObs} but for observables $R_{1c,1s,2c,2s,C,S}$.}\label{Tab:BinnedObsRi}
\begin{tabular}{ccccccccc}
\\ \hline \hline
~~~Observables~~~ & ~~~SM~~~ &~~~$\delta C_9^{\mu}$~~~ & ~~~$\delta C_{10}^{\mu}$~~~ & ~~~$\delta C_9^{\mu}= C_{10}^{\mu\prime}$~~~ & ~~~$\delta C_L^{\mu}$~~~& ~~~$\delta C_9^{\mu}=- C_9^{\mu\prime}$~~~ \\ \hline

$R_{1s}$ [1,4] GeV$^2$ & $0.982^{+0.002}_{-0.002}$ & $1.14^{+0.02}_{-0.03}$ & $0.84^{+0.03}_{-0.02}$ & $1.06^{+0.03}_{-0.03}$ & $0.94^{+0.03}_{-0.03}$&  $1.19^{+0.02}_{-0.03}$  \\

$R_{2s}$ [1,4] GeV$^2$ & $0.966^{+0.001}_{-0.001}$ & $1.12^{+0.01}_{-0.02}$ & $0.83^{+0.02}_{-0.02}$ & $1.04^{+0.02}_{-0.03}$ & $0.92^{+0.02}_{-0.03}$ &  $1.17^{+0.02}_{-0.02}$ \\

$R_{1c}$ [1,6] GeV$^2$ & $1.007^{+0.001}_{-0.001}$ & $0.83^{+0.02}_{-0.01}$ & $0.85^{+0.02}_{-0.02}$ & $0.77^{+0.02}_{-0.01}$ & $0.80^{+0.01}_{-0.01}$ & $0.68^{+0.06}_{-0.03}$  \\

$R_{2c}$ [1,6] GeV$^2$ & $0.977^{+0.001}_{-0.001}$ & $0.81^{+0.02}_{-0.01}$ & $0.83^{+0.02}_{-0.02}$ & $0.75^{+0.02}_{-0.01}$ & $0.78^{+0.01}_{-0.01}$ & $0.66^{+0.05}_{-0.03}$  \\

$R_{S}$ [1,4] GeV$^2$ & $0.985^{+0.003}_{-0.002}$ & $1.14^{+0.02}_{-0.03}$ & $0.84^{+0.03}_{-0.02}$ & $1.07^{+0.03}_{-0.03}$ & $0.95^{+0.03}_{-0.03}$ & $1.19^{+0.02}_{-0.03}$  \\

$R_{C}$ [1,6] GeV$^2$ & $0.9994^{+0.0007}_{-0.0008}$ & $0.82^{+0.02}_{-0.01}$ & $0.85^{+0.02}_{-0.02}$ & $0.77^{+0.02}_{-0.01}$ & $0.80^{+0.01}_{-0.01}$ & $0.68^{+0.06}_{-0.03}$
 \\ \hline \hline
\end{tabular}
\end{table}

Next, we study binned observables in proper ranges of $q^2$. As shown in Table~\ref{Tab:BinnedObsRi}, once $R_{1s}$ or $R_{2s}$ is measured with the same accuracy as that of theoretical predictions, the five different NP scenarios can be either confirmed or excluded at $2\sigma$ confidence level. Therefore, the binned observables $R_{1s}$ and $R_{2s}$ are useful to distinguish between different NP scenarios, though less constraining than the $q^2$ spectra.
We therefore conclude that the ratios $R_{1s}$ and $R_{2s}$ could uniquely distinguish between the five different NP solutions in the one-dimensional scenario if they can be measured with a precision of percent level.

On the experimental side, the angular coefficients $I_{1s}$ and $I_{2s}$ are not directly measured. Instead, the measured quantities are the branching ratio ${\rm BR}$ and the longitudinal
polarization $F_L$ as combinations of  $I_{1c}$, $I_{1s}$, $I_{2c}$, and $I_{2s}$, which read
\begin{eqnarray}
&&\frac{d{\rm BR}}{dq^2}=\frac{\tau_B}{2}\frac{d\left(\Gamma+\bar{\Gamma}\right)}{dq^2}=\frac{\tau_B}{4}\left[\left(3\Sigma_{1c}(q^2)-\Sigma_{2c}(q^2)\right)+2\left(3\Sigma_{1s}(q^2)-\Sigma_{2s}(q^2)\right)\right],\\
&&F_L=\frac{1}{2}\frac{3\Sigma_{1c}(q^2)-\Sigma_{2c}(q^2)}{d\left(\Gamma+\bar{\Gamma}\right)/dq^2},
\end{eqnarray}
where
\begin{eqnarray}
\Sigma_{1s,1c,2s,2c}(q^2)=I_{1s,1c,2s,2c}(q^2)+\bar{I}_{1s,1c,2s,2c}(q^2).
\end{eqnarray}
Clearly, the four angular coefficients $I_{1c}$, $I_{1s}$, $I_{2c}$, and $I_{2s}$ cannot be uniquely determined by two observables ${\rm BR}$ and $F_L$. However, we note that two combinations of angular coefficients $3\Sigma_{1c}-\Sigma_{2c}$ and $3\Sigma_{1s}-\Sigma_{2s}$ can be extracted from ${\rm BR}$ and $F_L$. Therefore,  we define two new ratios
\begin{eqnarray}
&&R_C(q^2)=\frac{3\Sigma_{1c}^\mu(q^2)-\Sigma_{2c}^\mu(q^2)}{3\Sigma_{1c}^e(q^2)-\Sigma_{2c}^e(q^2)},\\
&&R_S(q^2)=\frac{3\Sigma_{1s}^\mu(q^2)-\Sigma_{2s}^\mu(q^2)}{3\Sigma_{1s}^e(q^2)-\Sigma_{2s}^e(q^2)}.
\end{eqnarray}
It should be feasible to extract them experimentally. Therefore, we study the $q^2$ spectra and binned values of $R_{C,S}(q^2)$ in the SM and different NP scenarios. The relevant results are shown in Table~\ref{Tab:BinnedObsRi} and Fig.~\ref{Fig:Obsq2Ri}. It is clear that the ratio $R_S$ could uniquely distinguish between the five different NP solutions in the one-dimensional scenario if it can be measured with a precision of percent level.

\section{Summary and outlook}\label{Sec4}
In conclusion, in the one-dimensional NP case  only five scenarios, i.e., $\delta C_9^{\mu}$, $\delta C_{10}^{\mu}$, $\delta C_L^{\mu}$, $\delta C_9^{\mu}= C_{10}^{\mu\prime}$ and $\delta C_9^{\mu}=-C_9^{\mu\prime}$, can well describe  the latest LHCb, Belle, ATLAS, and CMS data. In order to discriminate the five different NP scenarios, we first ``constructed" four angular asymmetries $A_i$ and study the sensitivity of $A_i$ to new physics. We find that these observables $A_i$ cannot discriminate different NP scenarios. Further, we constructed a new ratio $R_S$ by revisiting a set of ratios $R_i$ and showed that the observable $R_S$ can uniquely discriminate  the five NP solutions in the one-dimensional case once they are measured in future experiments, by analysing the $q^2$ spectra of these observables and their binned results.

In the next few years, with the collection of more data at the LHCb,  Belle II experiments and improvement of experimental precision, such global fits should be  continually updated. In addition, new ideas should be explored to further reduce hadronic uncertainties. Furthermore, it will be  interesting to study the interplay between semi-leptonic baryon decays, such as $\Lambda_b\to\Lambda$, $\Xi_b\to\Xi$, and semi-leptonic meson decays. Because of rich helicity structures, if measured up to the same precision,  semi-leptonic baryon decays could constrain more tightly some relevant Wilson coefficients than their mesonic counterparts.

\section{Acknowledgments}
We thank Jorge Martin Camalich for  providing the code for the calculation of the $b\to s \ell\ell$ amplitudes.
This work is partly supported by the National Natural Science Foundation of China under Grants No.11735003, No.11975041, and No.11961141004, and the fundamental Research Funds for the Central Universities.

\section{Appendix}
\subsection{Discussion about NP in the electron and muon channels}
In our current work and most works performed so far,  NP is assumed to only originate from the muon mode. On the other hand, one can also assume that NP exists in the electron mode, or a combination of the electron and muon modes. In the electron mode, the signs of the fitted Wilson coefficients should be opposite to those of the muon mode if only the observables $R_K$ and $R_{K^*}$ are taken into account. The case of allowing for NP in the electron channel has been discussed quite extensively in Refs.~\cite{Ciuchini:2017mik,Hurth:2017hxg}~(for some recent studies see, e.g., Refs.~\cite{Kumar:2019qbv,Ciuchini:2019usw,Alok:2019ufo,Alguero:2018nvb,Alguero:2019pjc}). Here, we investigate a more general scenario that NP comes from a combination of electron and muon channels. As a result, more experimental data need to be taken into account, such as the branching fraction~\cite{Aaij:2013hha} as well as  eight angular observables~\cite{Aaij:2015dea,Aaij:2020umj} for the $B \to K^* e^+ e^-$ decay. Thus, the total number of data fitted becomes 103.
\begin{figure}[h!]
\centering
\begin{tabular}{ccc}
  \includegraphics[width=7cm]{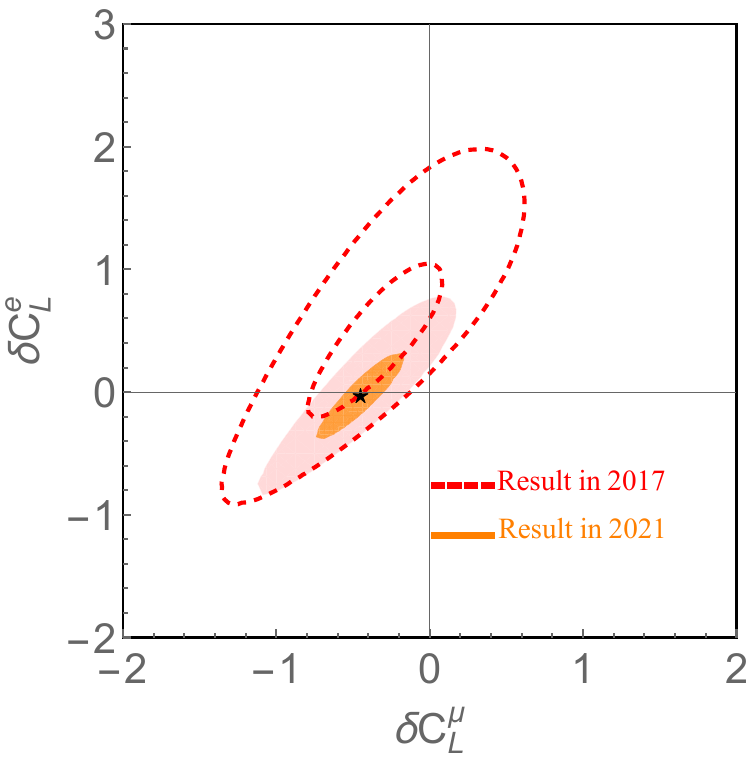}
& \hspace{0.5cm} &
 \includegraphics[width=7cm]{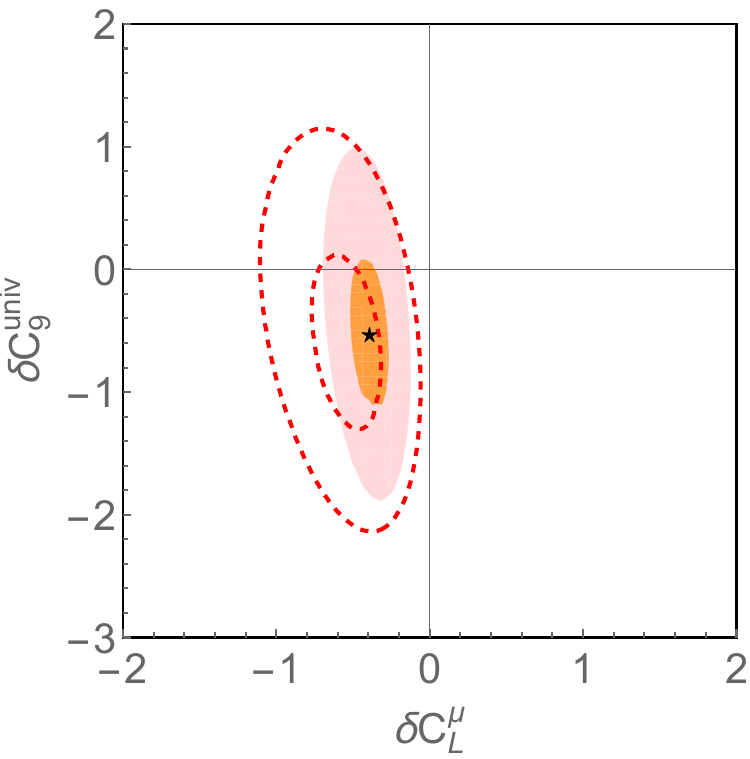}
\end{tabular}
\caption{Contours for the two-parameter fits ($\delta C_L^\mu$, $\delta C_L^e$) and ($\delta C_L^\mu$, $\delta C_9^{\rm univ}$) including the full data set~(103 data). The regions in orange and light-red represent the 1$\sigma$ and 3$\sigma$ bounds. The lines in red correspond to those in Fig.7 of Ref.~\cite{Geng:2017svp}.\label{Chp3FigelecCase}}
\end{figure}

In Fig.~\ref{Chp3FigelecCase}, we show the two-parameter fits for the ($\delta C_L^\mu$, $\delta C_L^e$) and ($\delta C_L^\mu$, $\delta C_9^{\rm univ}$) cases including the full 103 data set. Here, one requires $\delta C_9^{\rm univ}=\delta C_9^\mu =\delta C_9^e$. Compared to the results in 2017, all the NP scenarios are better constrained as shown by the smaller contours. In addition, the pure $\delta C_L^{\rm univ}$ scenario does not change significantly and is still in agreement with the SM within the $1\sigma$ confidence level while other scenarios are not. More specifically, the significance of the SM exclusion in the pure $\delta C_L^\mu$ scenario becomes larger. Also, the new data favor a zero $\delta C_9^e$ contribution. It should be mentioned that these variations from 2017 to 2021 are mainly caused by the most precise $R_K$ from the LHCb experiment. The  results indicate that new physics is most likely to exist in the muon mode.

\subsection{$q^2$ spectra of $R_{3-9}$}
In Fig. 5, we show the $q^2$ spectra of $R_{3-9}$. Clearly, these ratios cannot discriminate the five NP scenarios.
\begin{figure}[h!]
\begin{tabular}{cc}
  \includegraphics[width=7cm,height=4.5cm]{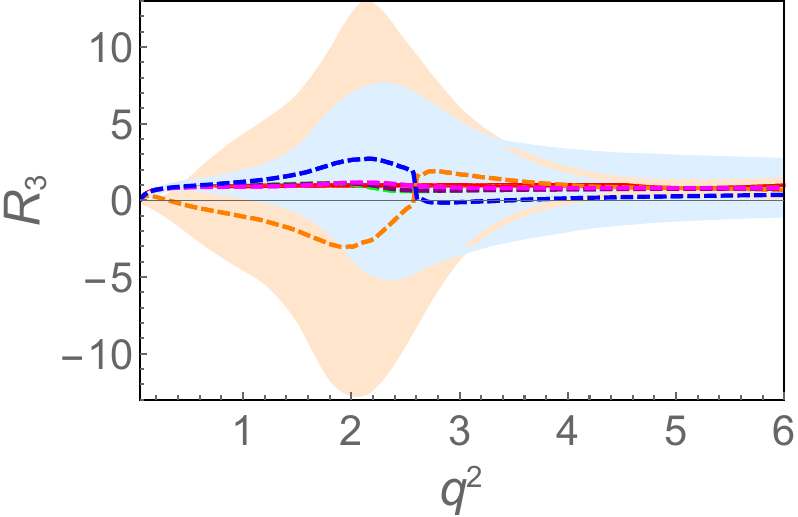}\\
  \includegraphics[width=7cm,height=4.5cm]{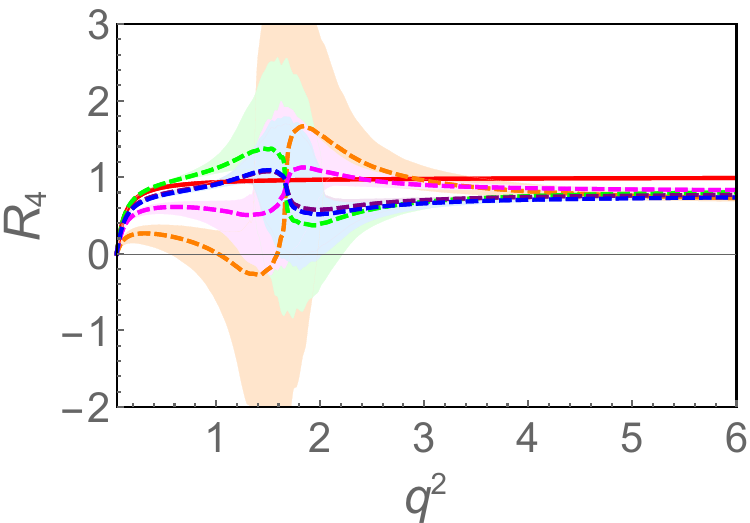}~~~\includegraphics[width=7cm,height=4.5cm]{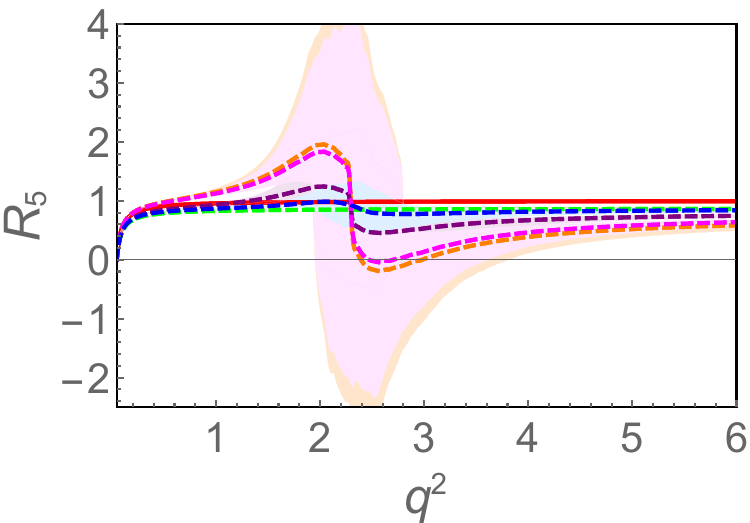}\\
  \includegraphics[width=7cm,height=4.5cm]{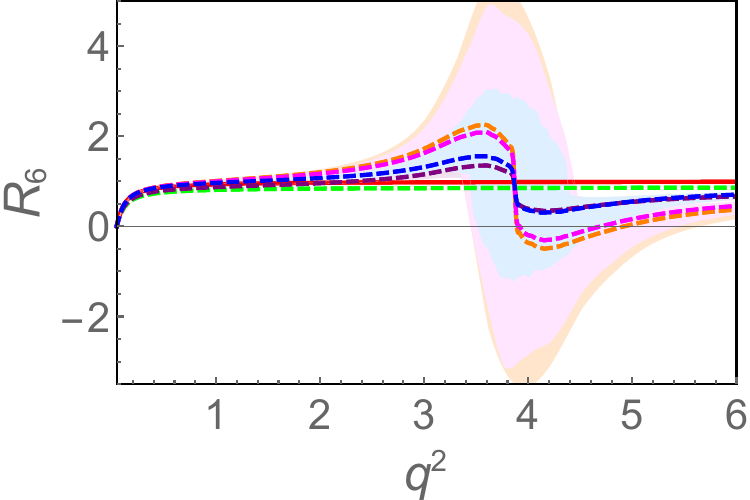}~~~\includegraphics[width=7cm,height=4.5cm]{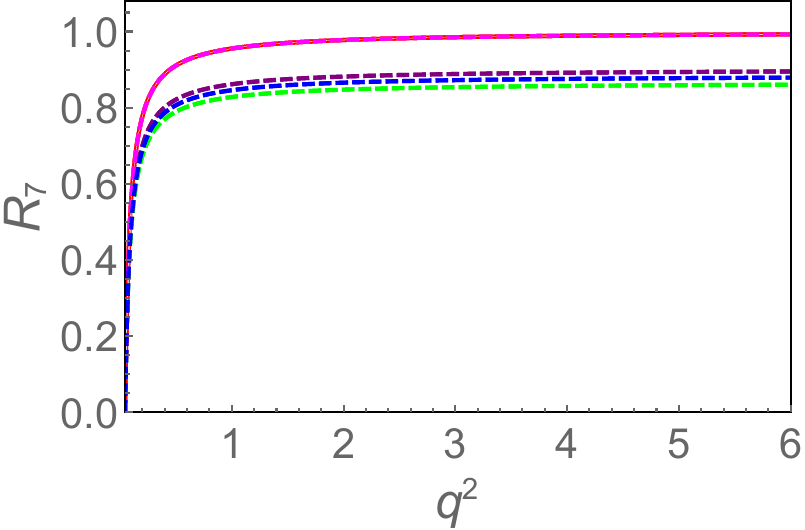}\\
  \includegraphics[width=7cm,height=4.5cm]{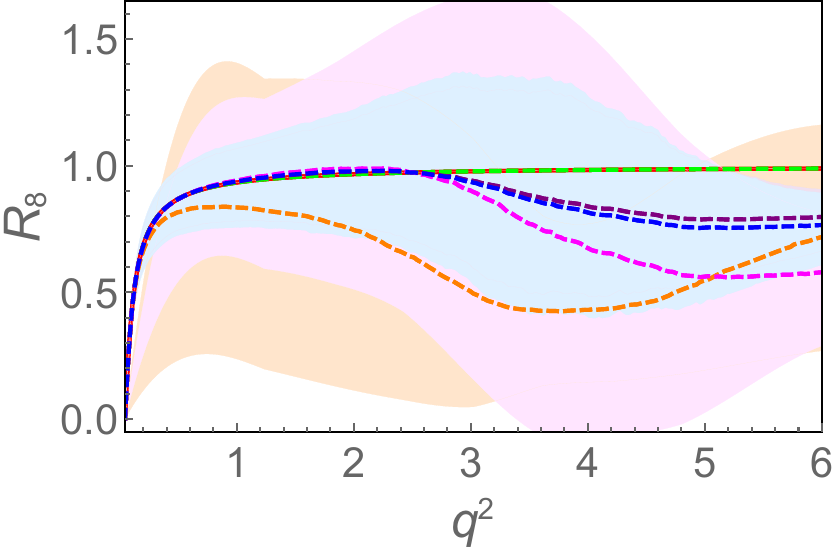}~~~\includegraphics[width=7cm,height=4.5cm]{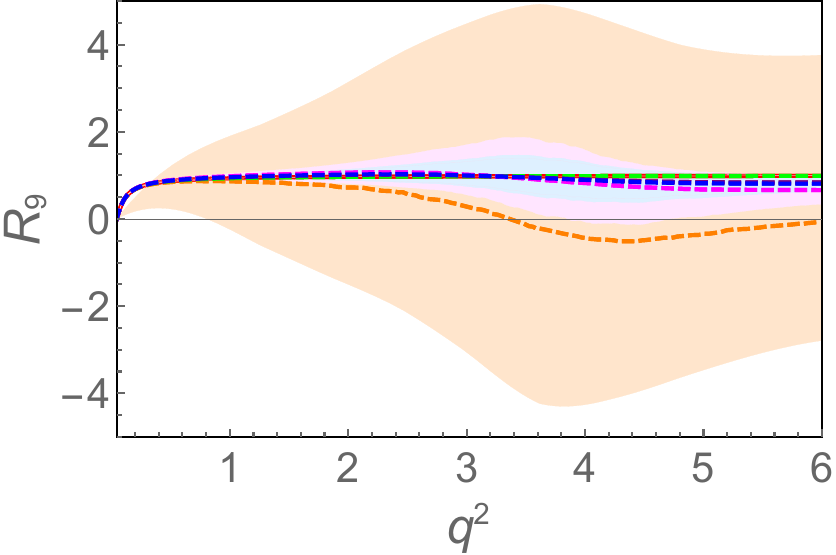}
\end{tabular}
\caption{Same as Fig.~\ref{Fig:Obsq2Ri} but for observables $R_{3-9}$.}\label{Fig:Obsq2Rirest}
\end{figure}

\bibliography{RKs}
\end{document}